\documentstyle[12pt,epsfig]{article}                        
\newcommand{\ber}{\begin{eqnarray}}
\newcommand{\eer}{\end{eqnarray}}
\newcommand{\bea}{\begin{equation}}
\newcommand{\eea}{\end{equation}}
\newcommand{\del}{\partial}
\begin{document}
\title{\bf Computing quantum correlation functions by Importance Sampling method based on  path integrals\\} 
\author{\bf Sumita Datta$^{1,2}$\\
$^1$ Allinace School of Applied Mathematics, Alliance University,\\ Bengaluru 562 106, India\\ 
$^2$ Department of Physics, University of Texas at Arlington,\\Texas 76019, USA\\}  
\maketitle
\begin{abstract}
An importance sampling method based on Generalized Feynman-Kac method has been used to calculate the mean values of quantum observables from quantum correlation functions for many body systems both at zero and finite temperature. Specifically, 
the expectation values $<r_{i}^n>$, $<r_{ij}^n>$, $<r_{i}^{-n}>$ and $<r_{ij}^{-n}>$ for the ground state of the lithium and beryllium and 
 the density matrix, the partition function, the internal energy and the specific heat of a system of quantum harmonic oscillators are computed, in good agreement with the best nonrelativistic values for these quantities. Although the initial results are encouarging, more experimentation will be needed to improve the other existing numerical results beyond chemical accuracies specially for the last two properties for lithium and beryllium. Also more work needs to be done to improve the trial functions for finite temperature calculations. 
\end{abstract}
\newpage 
\section{Introduction}
Feynman-Kac path integral method[1,2,3] provides a very accurate way of calculaing energy and other properties of quantum many body systems for the ground state i.e., for temperature $T=0$. Since this method makes a connection between the statistical mechanics and quantum theory, this formalism can also be applied to calculate the thermodynamic or finite temperature properties of quantum many body systems.
Our goal is to calculate zero and the finite temperature properties of many body systems using path integral technique[4,5,6,7,8]. As a matter of fact we can use the same algorithm for the zero and finite temperature properties.  

First we discuss the zero temperature properties. In Ref 5 Cafferel and Claverie point out that the original Feynman-Kac method[2,3] suffers from slow convergence rate because its underlying diffusion process is non-ergodic[9]. When reformulated to include a reference function associated with the correct sysmmetry of the state, the resulting diffusion process converges much more quickly. Using this Generalized Feynman-Kac method Caffarel and Claverie obtained the total energy and a few properties of some atomic and molecular systems to chemical accuracy(2-3 significant figures). In Ref[7] we used essentially this same method to calculate the ground state energies of the lithium and beryllium atoms. For our reference functions  we uesd the fully antisymmetric, explicitly correlated functions computed by Alexander and Coldwell[10]. By combining these high quality trial wavefunctions, which capture a large percentage of the correlation energy, with the Generalized Feynman-Kac method we recovered 100 \% of the total energy
for these systems with a low statistical error.

In this paper we examine whether this same method  and the same trial wavefunctions can be used to calculate the expectation values
$<r_{i}^n>$, $<r_{ij}^n>$, $<r_{i}^{-n}>$ and $<r_{ij}^{-n}>$ for the ground state of the lithium and beryllium.
As in Ref[7] our goal is to use the Generalized Feynman-Kac method to improve the value of these properties beyond that available from a
pure Variational Monte Carlo calculations. Unless otherwise indicated all the calculated values in this paper are in atomic units.

Since the imaginary time path integral formalism is equivalent to partion function, we use the same Generalized Feynman-Kac path integral method to calculate the statistical properties of a system of noninteracting harmonic oscillators. In the past, the thermodynamic propeties have been evaluated using path integrals[11,12,13] and by
stochastic method[14]. In ref[13] we had calculated the thermodynamic properties of a trapped Bose gas at low but finite temperature. Here in this paper, we consider a system of harmonic oscillators and calculate the Partition Function, Average energy, Density and Specific Heat for this system using the Quantum Monte Carlo method based on Generalized Feynman-Kac path Integral method. Now in ref[15] the quantum vs thermal fluctuations and their experimental implications have been addressed in relevance with an interacting harmonic chain. Apart from the theroretical significance of the problem, the interesting numerical aspect of it has motivated us to investigate on the noninteracting harmonic oscillators as it can act as a testing  bed for the problem addressed in ref[15] and  we aspire to do it in the future as an extension of the present endevour.

The organization of the contents in the article is as follows. In section 1 we introduce the problem and describe the organization of the paper. In section 2.1  we describe the path integral formalism used in this paper. In section 2.2 we rederive the Quantum correlation function in terms of stationary measure assocaited with the Generalized Feynman-Kac formalism. In section 3.1 we justify how using the same algorithm we can calculate zero and finite temperature properties. Section 3.2 contains the expression of all the above thermodynamical quantities using Ornstein-Uhlenbech process(a stochastic process adopted to accelerate the convergence). In section 4, we include the discussion of our results on zero and finite temperature properties. We conclude in section 5.     
In Table 1 and 2 we represent the raw simulation data for the expectation values of different quantities for lithium and beryllium respectively whereas in Table 3 and 4  we summarize the results for lithium and beryllium. In Table 5 we present the thermodynamical quantities for a single harmonic oscillator. In Table 6 we display the temperature variation of internal energy for a system of 10 independent harmonic oscillators interacting only through the heat bath. In Table 7, we explain our notations. 
\newpage
\section{Theory}
\subsection{Path integral formalism}
Metropolis and Ulam[16] were the first to exploit relationship between the Schr\"{o}dinger equation for imaginary time and the random-walk solution of diffusion equation.
Let us consider the initial value problem
\ber
& & \frac{\del{\psi}(x,t)}{\del{t}}=(\frac{\Delta}{2}-V){\psi}(x,t)\nonumber \\
& & {\psi}(x,0)=g(x)
\eer
where $x\in R^d$. 
The solution of the above equation can be written in Feynaman-Kac
representation as
\bea
{\psi}(x,t)=E_{x}[e^{-\int_{0}^{t}V(X(s))ds}g(X(t))]
\eea
for $V\in K_{\nu}$, the Kato class of potential[17].
A direct benefit of having the above representation is to recover the lowest energy eigenvalue of the Hamiltonian
$H=-\frac{\Delta}{2}+V $ for a given symmetry by applying the large deviation principle of Donsker and Varadhan as follows[18]:
\bea
{\lambda}_1=-\lim_{t\rightarrow \infty}\frac{1}{t}lnE_x[{e^{-\int_{0}^{t}V(X(s))ds}g(X(t))}]
\eea
The above
representation(Eq[2]) suffers from poor convergence rate as the underlying diffusion process-Brownian motion(Wiener Process[19]) is non-recurrent. So it is necessary to use a representation which employs a diffusion which unlike Brownian motion, has a stationary distributions.
For any twice differentiable $\phi(x)>0$ define a new potential U which is a
perturbation of the potential V as follows[6,7,8].
\bea
U(x)=V(x)-\frac{1}{2} \frac{\Delta \phi(x)}{\phi(x)}\,.
\eea
Also consider
\bea
\nu(x,t)=E_{x}[e^{-\int_{0}^{t}U(Y(s))ds}h(Y(t))]
\eea
which is Feynman-Kac solution to 
\ber
\frac{\del\nu(x,t)}{\del t}
& & =\frac{1}{2}\Delta \nu(x,t)+\frac{\nabla \phi(x)}{\phi (x)}\nabla \nu(x,t)-U(x)\nu(x,t)\\
 & & =-L\nu(x,t)
\nonumber\\
& & \nu(x,0)=h(x)\,,
\nonumber\\
\eer
where $h$ is the initial value of $\nu(x,t)$.
The new diffusion $ Y(t) $ has an infinitesimal generator $A=\frac{\Delta}{2}+\frac{\nabla \phi}{\phi}\nabla $, whose adjoint is $A^{\star}(\cdot)=\frac{\Delta}{2}-\nabla(\frac{\nabla \phi}{\phi}(\cdot))$. Here ${\phi}^2(x)$ is a stationary density of $ Y(t) $, or equivalently, $A^{\star}({\phi}^2)=0$.
To see the connection between $ \nu(x,t) $ and $ \psi(x,t) $, observe that for $g=1 $ and $h=1$,
\bea
\nu(x,t)=\frac{{\psi}(x,t)}{\phi(x)}\,,
\label{former_23}
\eea
because $ \nu(x,t) $ satisfies Equation (6). The diffusion $ Y(t) $ solves the following stochastic differential equation:
$dY(t)=\frac{\nabla \phi(Y(t))}{\phi(Y(t))}+dX(t)$\,.
In GFK we generate a diffusion process with the aid of a twice differentiable non-negative function  ${\phi}_0(x)$ by considering the following Hamiltonian:
\bea
H_0=-\frac{\Delta}{2}+U_0
\eea
\bea
U_0=e_0+\frac{1}{2} \frac{\Delta {\phi}_0(x)}{{\phi}_0(x)}
\eea
Here ${\phi}_0$ has been chosen as a trial function associated with the symmetry of the problem.
The main reason for introducing $H_0$ with a localized diffusion process is that it possesses a stationary distribution, unlike the nonlocalized Brownian motion process that escapes to infinity. This way one achieves the so called important sampling goal in the actual numerical computation, and thereby reduces the computing time considerably.
Now let us decompose the Hamiltonian of the quantum mechanical problem into two parts: $H=H_0+V_p$
In terms of the energy associated with the trial function,$e_0$, we now define a new perturbed potential  
\bea
V_p(x)=V-U_0
=V-e_0-\frac{1}{2} \frac{\Delta {\phi}_0(x)}{{\phi}_0(x)}
\eea
\newpage
\subsection{Correlation functions in Generalized Feynman-Kac represenation}
In this section we will discuss how to get the mean values of simple operators $ A_1, A_2......A_n$  from correlation 
functions. Let $g,h\in L^2$(g and h are the initial value of the trial fuctions as described in section 2.1) and $H=H_0+V_p$ and define a quantity 'Correlation function' in a statistical sense in imaginary time represenatation as follows[20]: 
\ber 
& & C(t_1,t_2,.............t_n)=E(A_{t_1},A_{t_2}.....A_{t_n})\nonumber\\
& &=\lim_{n\rightarrow \infty}\int \prod_{j=1}^{q}\int \prod_{k=1}^{n}{dx_1}^jA(x_1).........A(x_q)g(x(t_0))h(x(t_n))
<x_k|(e^{P/n}e^{Q/n})^n|x_{k-1}>\nonumber\\
\eer
Here $A_t=F(X_t)$ is a random variaible in the sense of probability theory and ${X_t}$ is the Markov process canonically associated with the Hamiltonian.
Now if $\hat{A}+\hat{B}=(t_k-t_{k-1})H$, then
\ber
\lim_{n\rightarrow \infty}(e^{P/n}e^{Q/n})^n=e^{-(t_k-t_{k-1})H}
\eer
Using the expression(Trotter Product) Eq(12) can be written as  
\ber
& & E(A_{t_1},A_{t_2}.....A_{t_n})\nonumber\\
&&=<g|e^{-(t_1-t_0)(H-e_0)}A_1e^{-(t_2-t_1)(H-e_0)}A_2.........A_qe^{-(t_f-t_q)(H-e_0)}|h>\nonumber\\
& & <g|e^{-(t_1-t_0)(H-e_0)}A(x_1)e^{-(t_2-t_1)(H-e_0)}A(x_2).........A(x_q)e^{-(t_f-t_q)(H-e_0)}|h>\nonumber\\
&&=\int_{\Omega(t_0,t_f)}g(X(t_0))A(X(t_1))A(X(t_2))......A(X(t_q))h(X(t_f))e^{-\int{V}_p(X(s))ds}D^{\phi}_0 X \nonumber\\
\eer
In the above expression we have considered a Markov process $X(t)$ and have defined $X(t_1)=x_1$, $X(t_2)=x_2$........$X(t_n)=x_n$.
Now taking $A_{t_1}=A$ and $A_{t_2}=A_{t_3}=.......A_{t_q}=1$, one can write
\bea
C(t;t_1)=<g{\phi}_0|e^{-(t_1-t_0)(H-e_0)}Ae^{-(t_2-t_1)(H-e_0)}|h{\phi}_0>
\eea
Now following ref[5], one can show that
the expectation value of the operator $A$ with respect to trial function for ith symmetry state, ${\phi}_{i0}$ is given by
\bea
{\langle {\phi}_{i0} |A|{\phi}_{i0} \rangle}
=\lim_{t\rightarrow \infty} \frac{C(t;t_1)}{C(t)} \forall t \in] t_0,t_f [
\eea
\ber
{\langle {\phi}_{i0} |A|{\phi}_{i0} \rangle}
& & =\lim_{t\rightarrow \infty} \frac{\int_{\Omega(t_0,t_f)}g(X(t_0))A(X(t_1))h(X(t_f))e^{-\int{V}_p(X(s))ds}D^{\phi}_0 X }{\int_{\Omega(t_0,t_f)}f(X(t_0))g(X(t_f))e^{-\int{V}_p(X(s))ds}D^{\phi}_0 X }
\eer
For $g=h=1$, $t_0=0$ and $t_f=t$ and $A=A(X(t))$
\ber
{\langle {\phi}_{i0} |A|{\phi}_{i0} \rangle}
& & =\lim_{t\rightarrow \infty} \frac{\int_{\Omega(0,t)}A(X(t_1))e^{-\int{V}_p(X(s))ds}D^{\phi}_0 X }{\int_{\Omega(0,t)}e^{-\int{V}_p(X(s))ds}D^{\phi}_0 X }
\eer
We calculate the expectation value of different operators employing the above formula. Now if the  
expectation value is calculated with respect to the ${\phi}_{i0}$ using Ornstein-Uhlenbeck process $Y(t)$ defined in the previous section thenone can write the above expectation value as 
\bea
{\langle Y|A|Y\rangle}=
\frac{\lim_{t\to\infty}\int dY(t)A(Y(t))e^{-\int{{V}_p(Y(s)ds}}}
{\int dY(t)e^{-\int{V}_p(Y(s)ds}}\,.
\label{former_13}
\eea
This is going to the key formula for calculating all the zero and finite temperature properties. 
\section{A Statistical approach to a system of Harmonic Oscillators}
Let us consider a system of noninteracting harmonic oscillators in thermal equilibrium. Each oscillator is independent and they interact only with the heat bath. The Partion function, Free Energy, Density, Average Energy and Specific Heat for the $ith$ oscillator can be defined as 
follows[21]:\\
Partition Function=$Z_i=\sum_ne^{-\epsilon_n^i/kT}$ where $ \epsilon_n^i=\hbar\omega_i(n+1/2)$\\
Free Energy=$-k_BTlnZ_i$\\
Average Energy of a single oscillator in thermal equilibrium=$\frac{1}{Z_i}\sum_n\epsilon_n^ie^{-\epsilon_n^i/kT}$\\
Specific heat at constant volume=$C_v=\frac{\del{U_i}}{\del{T}}=k_B{\beta}^2\frac{1}{Z_i(\beta)}[\sum_n{\epsilon_n^i}^2e^{-\epsilon_n^i/k_BT}-(\sum_n\epsilon_n^ie^{-\epsilon_n^i/k_BT})^2]$\\
Density=$Z_i^2$\\
Then for M oscillators,
$F=\sum_iF_i$, $U=\sum_iU_i$\\
In the section 3.2, the expressions for the above thermodynamic quantities have been derived in terms of a probabilistic measure called Wiener measure. While calculating those
numercally we actually evaluate them with respect to a stationary measure and find the expectation values with respect to Ornstein-Uhlenbeck process. All the thermodynamic quantities are calculated in $\hbar=\omega_i=k_B=1$, system of units, where $\hbar$ is the Planck's constant, $\omega_i$ is the frequency of ith oscillator and $k_B$ is the Boltzmann's constant.

\subsection{Finite Temperature Monte Carlo}
The temperature dependence comes from the realization that the imaginary time propagator is identical to the temperature dependent density matrix if $t=\beta=\frac{1}{T}$ holds.\\
This becomes obvious when we consider the equations[21]\\
\bea
-\frac{\del k(2,1)}{\del t_2}=H_2k(2,1)
\eea
and 
\bea
-\frac{\del \rho(2,1)}{\del \beta}=H_2\rho(2,1)
\eea
and compare
\bea
k(2,1)=\sum_{{i}^{\prime}}{\phi}_{{i}^{\prime}}(x_2){{\phi}_{{i}^{\prime}}}^{\star}(x_1)e^{-(t_2-t_1)E_{{i}^{\prime}}}
\eea
and
\bea
\rho{(2,1)}=\sum_{{i}^{\prime}}{\phi}_{{i}^{\prime}}(x_2){{\phi}_{{i}^{\prime}}}^{\star}(x_1)e^{-\beta E_{{i}^{\prime}}}
\eea
where $k(2,1)$ and ${\rho}_(2,1)$ are the imaginary time propagator and the temperature dependent density matrix respectively. 
For zero temperature Feynman-Kac procedure we had to extrapolate to $ t\rightarrow \infty $. For finite run time t in the simulation we have 
finite temperature results.
\subsection{Thermodynamic Quantities with probabilistic measure}
A particular temperature 'T' is said to be finite if $\Delta E < k_BT$ holds. Remember
for $T=0$
\bea
{\lambda}_1=-{\lambda}_T-\lim_{t\to\infty}\frac{1}{t}ln(E_{x_0}exp\{-\int_{0}^{t}V(X(s))ds\})
\eea
Using Wiener measure, we calculate the following thermodynamic quantities:
The partition function
\bea
Z(x,\beta)=\int_{\Omega}e^{-\int_{0}^{\beta}V(X(s))ds}d{\Omega}
\eea
The density matrix
\bea
\rho(\beta)= [\int_{\Omega}e^{-\int_{0}^{\beta}V(X(s))ds}d{\Omega}]^2
\eea
Again
Internal energy
\bea
U(\beta)=\frac{\int_{\Omega}x^2e^{-\int_{0}^{\beta}V(X(s))ds}d{\Omega}}{\int_{\Omega}e^{-\int_{0}^{\beta}V(X(s))ds}d{\Omega}} 
\eea
Now for finite T, the free energy
\bea
F(x,\beta)=\frac{ln\psi}{\beta}
\eea
where $\psi$ has been defined in Eq(2).
Specific heat
\bea
C(\beta)={\beta}^2\frac{\int_{\Omega}x^4e^{-\int_{0}^{\beta}V(X(s))ds}d{\Omega}-(\int_{\Omega}x^2e^{-\int_{0}^{\beta}V(X(s))ds}d{\Omega})^2}{\int_{\Omega}e^{-\int_{0}^{\beta}V(X(s))ds}d{\Omega}}
\eea
Here in this section we define the above thermodynamic quantities in terms of Brownian process, {X(t)}. But for actual computation of the expectation values for different operators we use Eq(19) which involves an  Ornstein-Uhlenbeck Process, $Y(t)$ with a stationary measure. In the case of a harmonic oscillator for the above quantities we have exact analytical expressions. For a Browmian motion which starts at zero, the Partition function, Internal energy and the Specific heat can be calculated using the one dimensional Cameron-Martin formula[22,23].
Partition Function
\bea
Z=1/\sqrt{\cosh{t}}
\eea 
Internal Energy
\bea
U=(1/2)\tanh{t}
\eea
Specific heat
\bea
C=0.5t^2(1/{\cosh}^2{t})
\eea
\section{Results and discussions} 
\subsection{Calculation of properties at T=0}
In a path integral calcultions each path consists of a sequence  of steps. In our
program the stepsize is fixed and the direction of the path is chosen randomly.
For each system a number of paths are generated with a specific path length. These are then summed to produce an average value and a statistical error. In order to examine the behavior of our properties as a function of path length, we compute several different path lengths - from 8 to 80 units of time. The averaged values of our properties at each of these path lengths are given in Tables 1 and 2 for the lithium and beryllium respectively.\\
For the lithium ground state we use the wavefunctions form
\bea
\psi=A [(r_{3}-c)e^{hyll-\alpha r_{1}-\beta r_{2}-\gamma r_{3}}] \\
\eea
where A is the antisymmetrization operator and
\bea
hyll=\sum_{k=0}a_{k}{q_{1}}^{\hat{c}}{q_{2}}^{\hat{d}}{q_{3}}^{\hat{e}}{q_{12}}^{\hat{f}}{q_{13}}^{\hat{g}}{q_{23}}^{\hat{h}}
\eea
and $q_{i}=\frac{r_{i}}{(1+br_{i})}$,and $q_{ij}=\frac{r_{ij}}{(1+br_{ij})}$
Here,${\hat{c}},{\hat{d}},{\hat{e}},{\hat{f}},{\hat{g}}$ and ${\hat{h}}$ are integers(0,1,....)that have been preselected for each
value of k. In all, this trial wavefunction contains 88 adjutable parameters.
As can be seen from Eq(24) the most accurate estimate of the energy is obtained when we extrapolate our results to infinite time. We do this by performing a least square fit to the functional form $E(t)=E(infinity)+a/t$. We have verified that the 10 path lengths selected with runtime selected from 8 to 80 are more than enough to perform an accurate fit. Oher tests have confirmed that a stepsize of 1/30 has little influence on the value of extrapolated energy. The final value obtained for the energy , -7.478069(6) is in excellent agreement with the best nonrelativistic value for this system[24]. It is, however only slightly better than the value obtained from the Variational Monte Carlo calculation. Increasing the number of configurations in the latter would lower the statistical error and help to more clearly distingusih between the two calculations.\\

Unlike the energy there is no need to extrapolate any of the properties to infinite time. As can be seen in Eq(19) these values for ${r_i}^n$ and
 ${r_{ij}}^n$ become independent of time after a short period of equilibration.
In Table 3 we show that the results for these properties at the largest time step are in excellent agreement with the best nonrealtivistic estimates for this system. This agreement can, however, be attributed to the quality of the original trial wavefunction since the Generalized Feynman-Kac properties and  Variational Monte Carlo properties are statistically identical. To distinguish between these two methods would require substantially longer calculations by both methods. Other tests have confirmed that the stepsize used has little influence on these values.\\

Unlike the properties discussed above,the expectation values ${r_i}^{-n}$ and
${r_{ij}}^{-n}$ do not converge to the correct results. We believe that this may be due to the simple nature of our path calculation. Since these properties are dominated by the region around the origin, it may be necessary to sample this region more often in order to obtain an accurate result.
For the beryllium ground state we use the trial wavefunction form
\bea
\psi=A[((r_3-d)(r_4-d)+c(x_3x_4+y_3y_4+z_3z_4))exp(hyll-\alpha r_1-\beta r_2-\gamma r_3-\delta r_4)]
\eea
\bea
hyll=\sum_{k=0}a_{k}{q_{1}}^{\hat{c}}{q_{2}}^{\hat{d}}{q_{3}}^{\hat{e}}{q_{4}}^{\hat{f}}{q_{12}}^{\hat{g}}{q_{13}}^{\hat{h}}
{q_{14}}^{\hat{m}}{q_{23}}^{\hat{n}}{q_{24}}^{\hat{o}}{q_{34}}^{\hat{p}}
\eea
Here $\hat{c},\hat{d},\hat{e},\hat{f},\hat{g},\hat{h},\hat{m},\hat{n},\hat{o}$ and $\hat{p}$ are integers(0,1,2.....) which have been preselected for each value of k. This trial wavefunction contains 72 adjustable parameters and does not capture as much of the correlation energy as its lithium counterpart. After extrapolation we obtain an energy of -14.66695(5) which is significant improvement over the Variational Monte Carlo results and is in better agreement with the best nonrelativistic value for this  system[7,25]. Like the energy, our values for $<{r_i}^n>$and ${r_{ij}}^n$ show a marked difference with the variational Monte Carlo properties. These results in Table 4 are statistically significant. In all cases the Feynman-Kac results are in better agreement with the best nonrelativistic estimates for this system.  
Like their lithium counterparts the expectation values $<{r_i}^{-n}>$ and $<{r_{ij}}^{-n}>$
do not converge to the correct results.
\subsection{Calculation of properties at finite temperature}
In Table 5 we present our results for the thermodynamic quantities e.g. partition function, density, internal energy, free energy and specific heat  for a single harmonic oscillator at $T=2$ and in Table 6 we have the internal energy for a system of 10 independent oscillators interactiong only through heat bath using the formula given in Eq(19) of section 2.2. At the moment we see some order of magnitude agreement of the Feynman-Kac results with the results coming from Cameron-Martin formula. We believe that we need have a better trial function for the simulation at higher temperature. 
In Fig 1, 2 and 3, we see the quantum and statistical fluctuation for  a single harmonic oscillator. Temperature is a measure of statistical fluctuation and Planck's constant $\hbar$ is a measure of quantum fluctuations. In Fig 1, corresponding to a very low temperature $T=0.125$, we see the fast quantum fluctuations around  $x=0$ which takes over the slow tharmal fluctuations. Fig 2, with an increase in temperature up to $T=0.25$ the slow thermal fuctuations creeps in on the top of the fast quantum fluctuations. In Fig 3, we see that as temperature increased to a value $T=1$ the thermal fluctuations have completely taken over the fast quantum fluctuations and the whole configuration has 
 moved away from $x=0$. In Fig 4, we show the comparison of ground state density plot for a single oscillator for the FK data and Cameron-Martin formula[26]. In Fig 5, we show the temperature variation of the internal energy for the above chain of 10 independent harmonic oscillators.   
\begin{figure}[h!]
\includegraphics[width=3.5in,angle=-90]{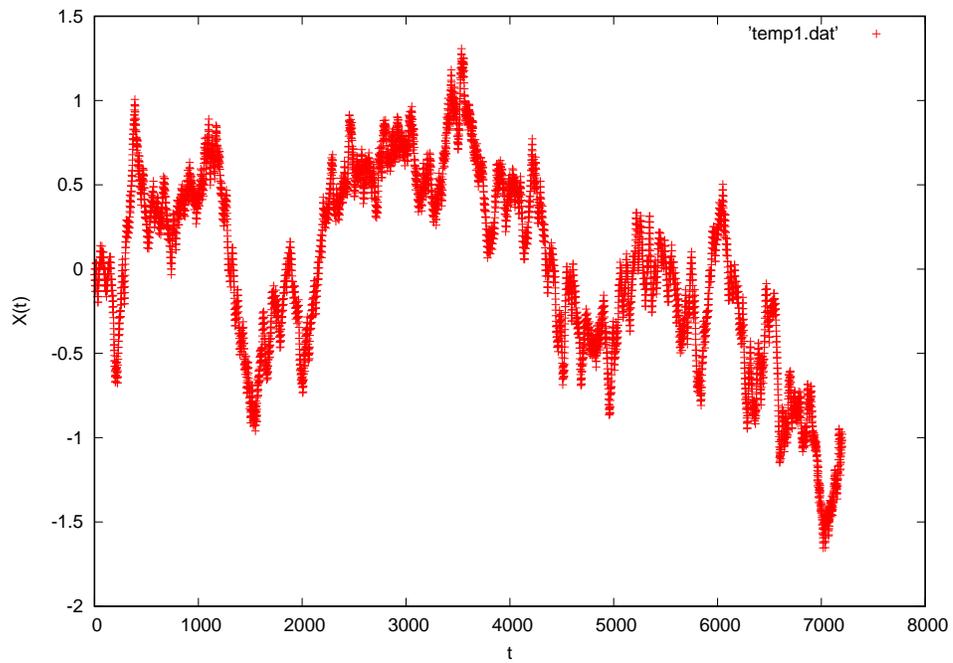}
\caption{A plot for the fluctuations at T=0.125}
\end{figure}
\begin{figure}[h!]
\includegraphics[width=3.5in,angle=-90]{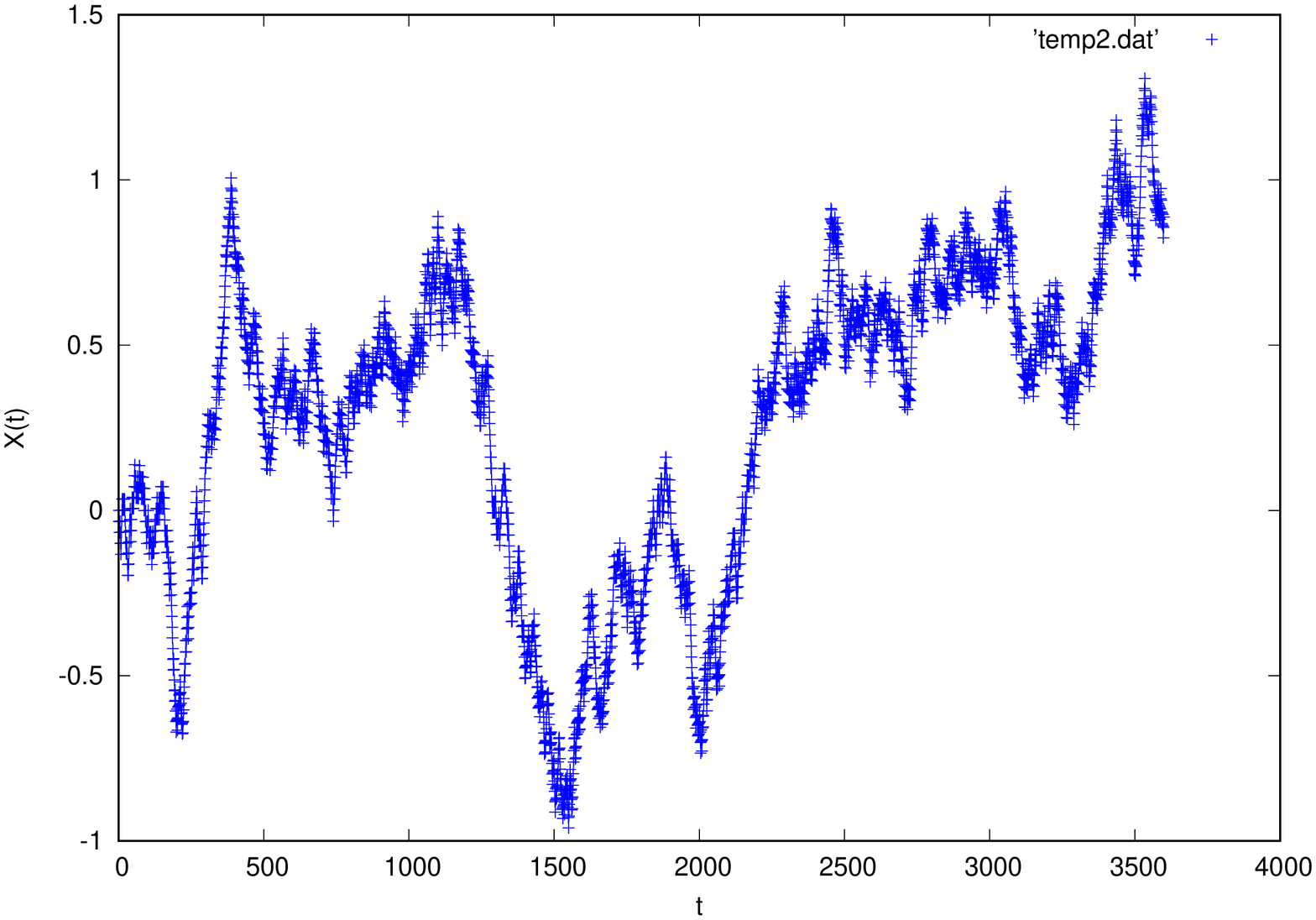}
\caption{A plot for the fluctuations at T=0.25}
\end{figure}
\newpage
\begin{figure}[h!]
\includegraphics[width=4in,angle=-90]{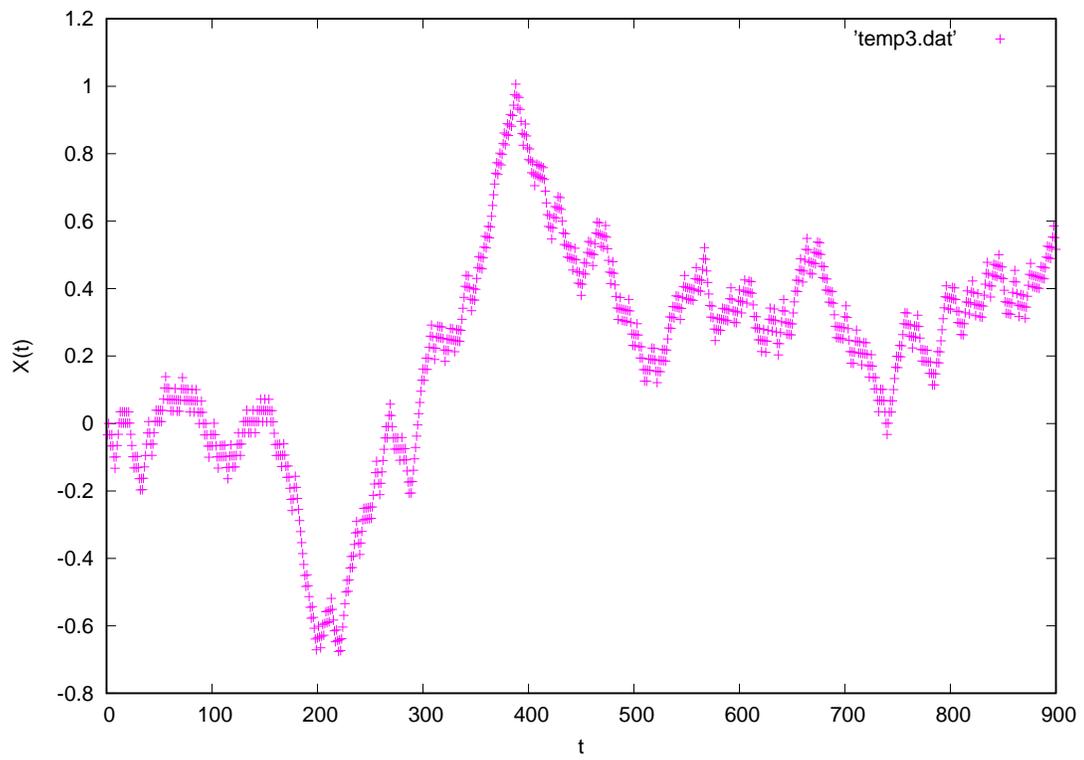}
\caption{A plot for the fluctuations at T=1}
\end{figure}
\begin{figure}[h!]
\includegraphics[width=4in,angle=-90]{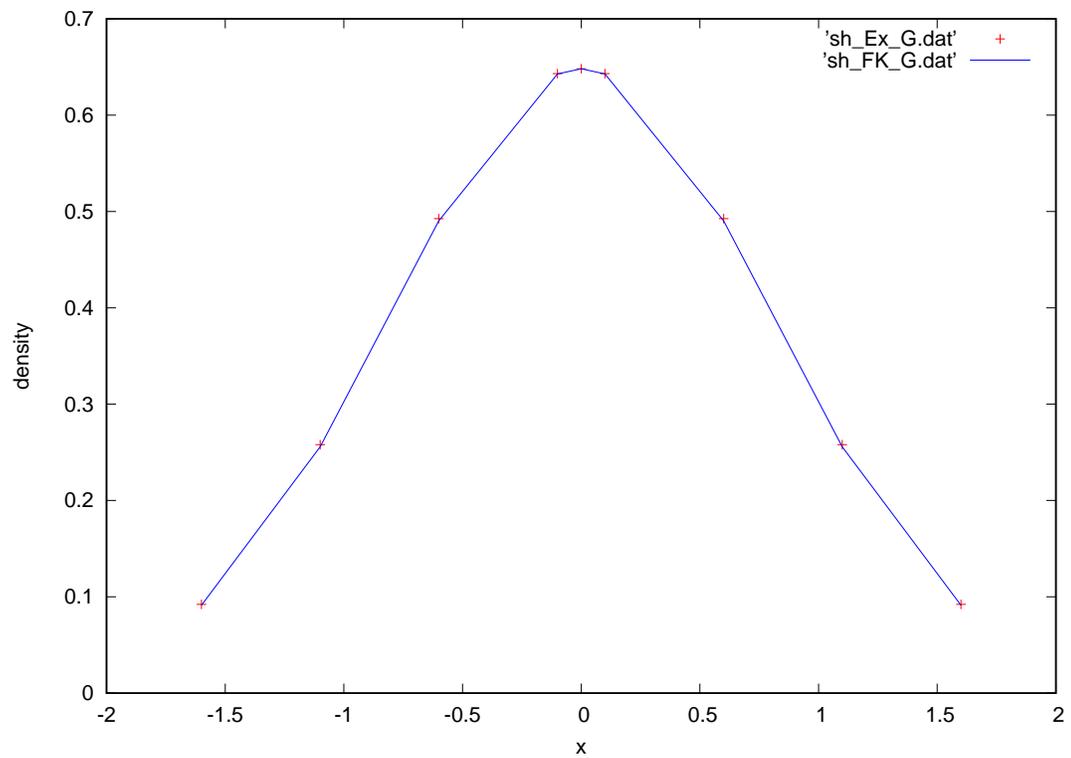}
\caption{A plot for the ground state density}
\end{figure}
\begin{figure}[h!]
\includegraphics[width=4in,angle=-90]{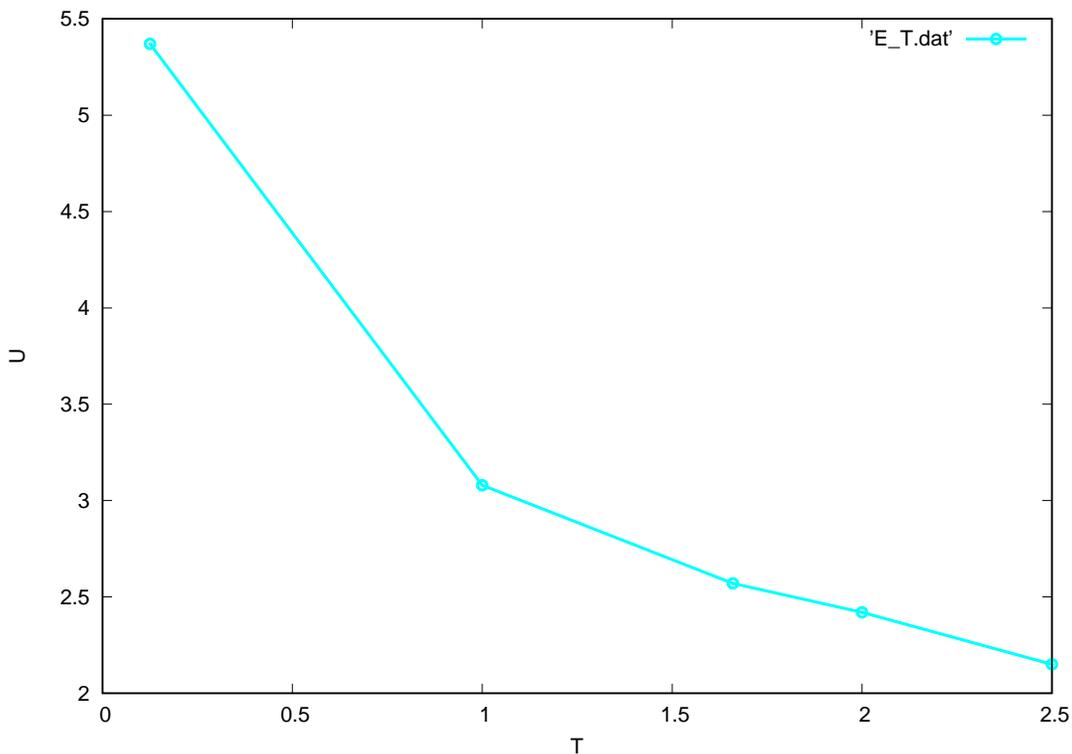}
\caption{A plot for the variation of internal energy with temperature}
\end{figure}
The data in Fig 5 are given by Table 6.

\section{Conclusions}
We have combined accurate trial wavefunctions with the Generalized Feynman-Kac
method to compute some properties of the ground state of the lithium  and beryllium  atoms and thermodynamic properties of a system of non-interacting harmonic oscillators in a temperture bath. As shown in Tables 3 and 4, our expectation values for  $<{r_i}^n>$ and $<{r_{ij}}^n>$ are in good agreement with the best nonrelativistic values for these systems. Because of the high quality of our trial wavefunctions the Feynmna-Kac method is able to begin near the correct result. Especially in the case of beryllium, however, we see that this method brings both the energy and the value for these properties closer to the correct nonrelativistic values. In contrast the Feynman-Kac method does not improve our results for $<{r_i}^{-n}>$ and $<{r_{ij}}^{-n}>$. We believe that this is due to the simple algorithm we used to generate the paths.

For the harmonic oscillator we can calculate the thermodynamic properties with a reasonably good agreement with the analytical results. Using the same algorithm one can calculate both zero and finite temperature  properties using Eq(19) in  section 2.2. For the details of the numerical implementation of the formaula for zero and finite temperature properties one can see the Appendix A of ref[27]. For the zero temperature we have been able to come up  with high quality trial functions and improve the variational Monte Carlo results. But for the non-intercating simple harmonic chain in heat bath we need to use better trial functions to guide our random walk so that we can achieve a better agreement with analytical results. At this moment we just have achieved the order of magnitude agreement between the FK results and the analytical results.
It would be interesting to test our numerical scheme for interacting harmonic chain and this work is currently underway.
In summary we feel that using the Generalized Feynman-Kac method to calculate the properties at zero and finite temperature is promising but clearly more work needs to be done.\\  
{\bf Acknowledgements:}
This work was partially supported by Department of Science and Technology, New Delhi, India, Grant no EMR/2016/005492).
\newpage
\begin{table}[h!]
\begin{center}
\caption {\bf Selected lithium results (in au)calculated using the Feynman-Kac path integral method at various times with a stepsize of $1/30$and 18550 paths. The number in the parentheses is the statiscal error.}
\begin{tabular}{ccccccc}
time & Energy & $<r_i> $ & $<r_i^2> $  & $<r_{ij}> $ & $ <r_{ij}^2> $\\
8 & -7.478927(28) & 4.623(12) & 14.83(8) & 7.945(22) & 29.81(17)\\
16 & -7.478519(18) & 4.867(13) & 17.00(10) & 8.415(25) & 34.18(21)\\
24 & -7.478380(14) & 4.961(14) & 17.97(11) & 8.604(27) & 36.07(24)\\
32 & -7.478294(12) & 4.995(14)& 18.33(12) & 8.669(28) & 36.79(24)\\
40 & -7.478244(10)& 5.017(15)& 18.59(13) & 8.717(28) & 37.31(25)\\
48 & -7.478215(9) & 4.978(15) & 18.24(12) & 8.640(28) & 36.59(25)\\
56 & -7.478194(9) & 4.987(16) & 18.51(22) & 8.667(29) & 37.20(45)\\
64 & -7.478176(8) & 4.985(15) &18.42(21) & 8.650(29) & 36.93(41)\\
72 & -7.478166(7) & 4.988(15) & 18.33(18) & 8.652(29) & 36.79(37)\\
80 & -7.478157(7) & 4.982(16) & 18.33(17) & 8.643(30) & 36.76(35)\\
time & $<r_i^{-1}> $ & $<r_i^{-2}> $ & $ <r_{ij}^{-1}> $\\
8  & 5.697(27) & 27.1(6) & 2.242(10)\\
16 & 5.619(28) & 27.1(8) & 2.211(10)\\
24 & 5.648(28) & 27.1(8) & 2.192(10)\\
32 & 5.632(28) & 27.1(12) & 2.202(13)\\
40 & 5.639(29) & 27.9(16) & 2.188(10)\\
48 & 5.661(29) & 27.6(10) & 2.189(10)\\
56 & 5.701(32) & 30.4(18) & 2.207(11)\\
64 & 6.615(28) & 26.1(7) & 2.187(10)\\
72 & 5.679(31) & 29.1(18) & 2.192(11)\\
80 & 5.599(28) & 25.6(6) & 2.212(12)\\
\end{tabular}
\end{center}
\end{table}
\newpage
\begin{table}[h!]
\begin{center}
\caption {\bf Selected beryllium results (in au)calculated using the Feynman-Kac path integral method at various times with a stepsize of $1/30$ and 7000 paths. The number in the parentheses is the statiscal error.}
\vskip 0.5cm
\begin{tabular}{ccccccc}
time & Energy & $<r_i> $ & $<r_i^2> $  & $<r_{ij}> $ & $ <r_{ij}^2> $\\
8  &-14.66946(14) & 5.889(25) & 15.47(12) & 15.03(7) & 50.45(41)\\
16 & -14.66822(11) & 6.012(31) & 16.26(14) & 15.34(8) & 52.81(48)\\
24 & -14.66782(10) & 6.011(36) & 16.37(16) & 15.35(10) & 53.09(53)\\
32 & -14.66762(9)  & 5.995(41) & 16.20(17) & 15.32(11) & 52.65(56)\\
40 & -14.66750(8) & 6.017(45) & 16.32(18) & 15.37(12) & 54.38(68)\\
48 & -14.66745(7) & 6.017(49) & 16.37(19) & 15.36(13) & 53.16(62)\\
56 & -14.66734(7) & 6.067(53) & 16.68(20) & 15.53(14) & 54.38(68)\\
64 & -14.66729(6) & 6.061(55) & 16.65(20) & 15.48(14) & 53.90(67)\\
72 & -14.66721(6) & 6.022(59) & 16.46(21) & 15.40(15) & 53.48(70)\\
80 & -14.66719(6) & 6.022(62) & 16.39(22) & 15.37(16) & 53.11(72)\\
time & $<r_i^{-1}> $ & $<r_i^{-2}> $ & $ <r_{ij}^{-1}> $\\
8  & 8.15(6) & 40.1(9) & 4.33(3)\\
16 & 8.11(70 & 40.0(9) & 4.27(3)\\
24 & 8.11(7) & 39.0(9) & 4.28(3)\\
32 & 8.18(8) & 40.8(10) & 4.29(4)\\
40 & 8.17(8) & 40.0(9) & 4.31(4)\\
48 & 8.08(9) & 39.9(10) & 4.31(4)\\
56 & 8.14(9) & 40.2(10) & 4.24(4)\\
64 & 8.14(10) & 40.1(10) & 4.28(5)\\
72 & 8.19(10) & 41.5(11)& 4.34(5)\\
80 & 8.15(10) & 40.6(10) & 4.32(5)\\
\end{tabular}
\end{center}
\end{table}
\begin{table}
\begin{center}
\caption {\bf Selected properties for the lithium atom (in au)computed using the Feynman-Kac path integral method with a stepsize of $1/30$  and 18550 paths. The VMC calculations were performed with 1024000 biased as random + mixed configurations. The number in the parentheses is the statistical error.}
\scalebox{0.9}{
\begin{tabular}{ccccc}
Property & Path Itegral & VMC & Literature\\
E & -7.478069(6) & 7.47800(3) & -7.4780603[7]\\
$<r_i> $  & 4.98(2) & 4.99(2) & 4.989523[7]\\
$<r_i^2> $ & 18.3(2) & 18.37(8)  & 18.354614[7]\\
$<r_i^{-1}> $ & 5.60(3) & 5.72(2) & 5.718109[7]\\
$<r_i^{-2}> $ & 25.6(6) & 30.1(1) & 30.21204[7]\\
$<r_{ij}> $ & 8.64(3) & 8.68(2) & 8.668396[7]\\
$ <r_{ij}^2> $ & 36.8(4) & 36.9(1) &36.848033[7]\\
$ <r_{ij}^{-1}> $ & 2.21(1) & 2.20(8)& 2.198211[7]\\
\end{tabular}}
\end{center}
\end{table}
\begin{table}[h!]
\begin{center}
\caption {\bf Selected properties for the beryllium atom (in au)computed using the Feynman-Kac path integral method with a stepsize of $1/30$  and 7000 paths. The VMC calculations were performed with 1024000 biased as random + mixed configurations. The number in the parentheses is the statiscal error.}
\begin{tabular}{ccccc}
Property & Path Itegral & VMC & Literature\\
E & -14.66695(5) & -14.6660(2) & -14.667355[8]\\
$<r_i> $  & 6.02(6) & 6.07(2) & 5.972388[8]\\
$<r_i^2> $ &16.4(2) & 16.84(5) & 16.24592[8]\\
$<r_i^{-1}> $&8.2(1) & 8.42(2) & 8.427348[8]\\
$<r_i^{-2}> $ & 41.0(1) & 57.7(2) & 57.5976[8]\\
$<r_{ij}> $ & 15.4(2) & 15.50(3) & 15.271674[8]\\
$ <r_{ij}^2> $ & 53.1(7) & 54.5(1) & 52.84896[8]\\
$ <r_{ij}^{-1}> $ & 4.32(5) & 4.35(1) & 4.374695[8]\\
\end{tabular}
\end{center}
\end{table}
\vspace{-5 cm}
\newpage
\begin{table}[h!]
\begin{center}
\caption{\bf Selected Thermodynamic properties of a harmonic oscillator computed using the Feynman-Kac path integral method at temperature $ T=2 $ with a stepsize of $1/30$  and 10000 paths.}
\begin{tabular}{cccc}

Property & Path Integral(FK) & Analytical\\
density $\rho$  &  0.840$\pm$ 0.0008   &  0.933\\
Partition function Z & 0.916$\pm$0.001 & 0.941\\
Free energy F & 0.1735$\pm$0.0002 & 0.120\\
Average energy U & 0.2424$\pm$ 0.004 & 0.231\\
Specific heat $C_s$ & 0.029$\pm$0.001 & 0.098\\
\end{tabular}
\end{center}
\end{table}
\begin{table}[h!]
\begin{center}
\caption{\bf Internal energy of a system of 10 independent harmonic oscillators interacting only through the heat bath computed using the Feynman-Kac path integral method at different temperature with a stepsize of $1/30$ and 10000 paths.}
\begin{tabular}{ccc}
Temperature & internal energy(Path Integral(FK) \\
2.5    &     2.15$\pm$0.003\\
2      &     2.42$\pm$0.004\\
1.66   &     2.57$\pm$0.004\\
1      &     3.08$\pm$0.008\\
0.125  &     5.37$\pm$0.003\\ 
\end{tabular}
\end{center}
\end{table}
\newpage
\begin{table}[h!]
\begin{center}
\caption{\bf Notation Table}
\begin{tabular}{ccc}
Notation/Phrase & Meaning \\
$r_i$      &   position vector of electrons inside the atom \\
$r_{ij}$   &   distance between two electrons\\
$X(t)$     &   Brownian motion with a non-ergodic probabilistic measure or\\ 
           &    Wiener Measure\\
$Y(t)$     &   A stochastic process with an ergodic or stationary measure\\     
$ \phi_0$  &   A trial function\\
$\phi_{i0}$ &  A trial function for ith state \\
\end{tabular}
\end{center}
\end{table}


\newpage
\begin{thebibliography}{99}
\bibitem{1} R. P. Feynman and A. R. Hibbs, Quantum Mechanics and Path Integrals(McGraw-Hill,NY(1965))
\bibitem{2} M. D. Donsker and M. Kac, J. Res. Natl. Bur. Stand {\bf 44},
581 (1950); see also, M.Kac, in Proceedings of the Second Berkeley Symposium
(Berkeley Press, California (1951)).
\bibitem{3} A. Korzeniowski, J.L. Fry, D. E. Orr and N. G. Fazleev, Phys Lett {\bf 69}, 893,1992
\bibitem{4} F. Soto-Equibar and P Claverie,in Stochastic Processes Applied to Physics and other A Rueda(World Scientific, Singapore,1983).
\bibitem{5} M.Cafferel and P. Claverie, J. Chem Phys. {\bf 88 }, 1088 (1988);{\bf 88}, 1100 (1988).
\bibitem{6} A. Korzeniowski, J Comp and App Math {\bf 66}, 333 (1996)
\bibitem{7} S. Datta, J. L Fry, N. G. Fazleev, S. A. Alexander and
R. L. Coldwell, Phys Rev A {\bf 61} R030502 (2000); S. Datta, Ph.~D
dissertation, The University of Texas at Arlington (1996).
\bibitem{8} S. Datta and J M Rejcek, Eur. Phys J. Plus{\bf 133} 202(2018)     
\bibitem{9} I. Karatzas, S. E. Shreve, Brownian Motion and Stochastic Calculus,(Springer-Verleg,NY, 1991)
\bibitem{10}S A. Alexander and R L. Coldwell, Int J Quantum Chem. {\bf 97},1001(1997)
\bibitem{11} M Cruetz and B Freedman, Ann. Phys {\bf 132},427(1981)
\bibitem{12} A Larson and F Ravndal, Am J Phys, {\bf 56},1129(1988)
\bibitem{13} S Datta, Int. J. Mod Phys., {\bf 22}, 4261(2008) 
\bibitem{14} C Huang, H Croger, X Q Luo and K J M Moriarty, Phys Lett A
{\bf 299}, 483(2002)
\bibitem{15} K. Sch\"{o}nhammer, Am J  Phys 82, 887 (2014) 
\bibitem{16} N. Metropolis and S. Ulam, J. Am. Stat. Assn. {\bf 44},335(1949).
\bibitem{17} T Kato, Commun. Pure Appl Math,{\bf 10},151(1957)
\bibitem{18} M. D. Donsker and S. R. Varadhan, in Proc. of the International
Conference on Function space Integration ( Oxford Univ. Press 1975)pp. 15-33.
\bibitem{19} N. Wiener, J. Math and Phys., {\bf 2},132,(1923).
\bibitem{20} G. Roepstorff, Path integral approach to quantum physics(Springer, 1994)
\bibitem{21} R. P. Feynman, Statistical Physics:a set of lectures (Reading Mass.: Addison-Wesley, 1998) 
\bibitem{22} A Korzeniowski, J Math Phys,{\bf 26}, 2189(1985)
{\bf 299}, 483(2002)
\bibitem{23} R. Durret, Brownian Motion and Martingales in Analysis(Wadsworth Mathematics Series,Belmont, CA,1984)
\bibitem{24} Z. C. Yan and G. W. F. Drake, Phys Rev A {\bf 52}, 3711(1995).
\bibitem{25} J Komasa, W Cenek and J Rychlewski, Phys Rev A {\bf 52}, 4500(1995)
\bibitem{26} A Korzeniowski, Probability in the Engineering Informational Sciences, {\bf 5}, 101, 1991
\bibitem{27} Sumita Datta, Vanja Dunjko and Maxim Olshanii, Physics,{\bf 4}, 12, (2022)
\end{thebibliography}
\end{document}